\documentclass[sigconf]{acmart}

\AtBeginDocument{%
  }

\copyrightyear{2026}
\acmYear{2026}
\setcopyright{cc}
\setcctype{by}
\acmConference[SIGIR '26]{Proceedings of the 49th International ACM SIGIR Conference on Research and Development in Information Retrieval}{July 20--24, 2026}{Melbourne, VIC, Australia}
\acmBooktitle{Proceedings of the 49th International ACM SIGIR Conference on Research and Development in Information Retrieval (SIGIR '26), July 20--24, 2026, Melbourne, VIC, Australia}
\acmDOI{10.1145/3805712.3809578}
\acmISBN{979-8-4007-2599-9/2026/07}

\usepackage{enumitem}
\usepackage{makecell}
\usepackage{xcolor}
\usepackage{cleveref}  
\usepackage{algorithm}
\usepackage{algpseudocode}
\usepackage{multirow}
\usepackage{pifont}
\usepackage{subcaption}
\usepackage{balance}
\newcommand{\cmark}{\ding{51}}
\newcommand{\xmark}{\ding{55}}  
\newcommand{\blue}[1]{\textcolor{black}{#1}}
\newcommand{\sigir}[1]{\textcolor{black}{#1}}

\makeatletter
\algrenewcommand\algorithmicrequire{\textbf{Input:}}
\algrenewcommand\algorithmicensure{\textbf{Output:}}
\makeatother

\begin{document}

\title[Decomposed Contextual Token Representations from Pretrained and Collaborative Signals]{Learning Decomposed Contextual Token Representations from Pretrained and Collaborative Signals for Generative Recommendation}


\author{Yifan Liu}
\authornote{Both authors contributed equally to this research.}
\affiliation{%
  \institution{University of Illinois Urbana-Champaign}
  \city{Champaign}
  \state{Illinois}
  \country{USA}
}\email{yifan40@illinois.edu}

\author{Yaokun Liu}
\authornotemark[1]
\affiliation{%
  \institution{University of Illinois Urbana-Champaign}
  \city{Champaign}
  \state{Illinois}
  \country{USA}
}\email{yaokunl2@illinois.edu}

\author{Zelin Li}
\affiliation{%
  \institution{University of Illinois Urbana-Champaign}
  \city{Champaign}
  \state{Illinois}
  \country{USA}
}
\email{zelin3@illinois.edu}

\author{Zhenrui Yue}
\affiliation{%
  \institution{University of Illinois Urbana-Champaign}
  \city{Champaign}
  \state{Illinois}
  \country{USA}
}
\email{zhenrui3@illinois.edu}

\author{Gyuseok Lee}
\affiliation{%
  \institution{University of Illinois Urbana-Champaign}
  \city{Champaign}
  \state{Illinois}
  \country{USA}
}
\email{gyuseok2@illinois.edu}

\author{Ruichen Yao}
\affiliation{%
  \institution{University of Illinois Urbana-Champaign}
  \city{Champaign}
  \state{Illinois}
  \country{USA}
}
\email{ryao8@illinois.edu}

\author{Yang Zhang}
\affiliation{%
  \institution{Miami University}
  \city{Miami}
  \state{Florida}
  \country{USA}
}
\email{zhang981@miamioh.edu}

\author{Dong Wang}
\affiliation{%
  \institution{University of Illinois Urbana-Champaign}
  \city{Champaign}
  \state{Illinois}
  \country{USA}
}
\email{dwang24@illinois.edu}
\renewcommand{\shortauthors}{Yifan Liu et al.}

\begin{abstract}
Recent advances in generative recommenders adopt a two-stage paradigm: items are first tokenized into semantic IDs using a pretrained tokenizer, and then large language models (LLMs) are trained to generate the next item via sequence-to-sequence modeling.
However, these two stages are optimized for different objectives: semantic reconstruction during tokenizer pretraining versus user interaction modeling during recommender training. 
This objective misalignment leads to two key limitations: (i) suboptimal static tokenization, where fixed token assignments fail to reflect diverse usage contexts; and (ii) discarded pretrained semantics, where pretrained knowledge—typically from language model embeddings—is overwritten during recommender training on user interactions. 
To address these limitations, we propose to learn \underline{DE}composed \underline{CO}ntextual Token \underline{R}epresentations (DECOR), a unified framework that preserves pretrained semantics while enhancing the adaptability of token embeddings.
DECOR introduces contextualized token composition to refine token embeddings based on user interaction context, and decomposed embedding fusion that integrates pretrained codebook embeddings with newly learned collaborative embeddings. 
Experiments on three real-world datasets demonstrate that DECOR consistently outperforms state-of-the-art baselines in recommendation performance.
Our code is available at~\footnote{https://github.com/yliuaa/DECOR.git}.
\end{abstract}





\maketitle

\section{Introduction}
Recent advances in Large Language Models (LLMs) have led to the emergence of \textit{generative recommendation} as a new paradigm for sequential recommendation~\cite{tiger,genrecsys_review}. Numerous efforts investigating generative recommenders have formulated recommendations as an auto-regressive sequence generation task, aligning naturally with the strengths of LLMs in modeling long-range dependencies and generating coherent sequences~\cite{letter,LCRec,etegrec,genrecsys_review}. \blue{Additionally, generative recommender systems naturally possess the advantage of handling cold-start items by leveraging pre-training knowledge to provide semantic priors and contextual understanding.}
Specifically, generative recommenders typically follow a two-stage training pipeline: \textit{item tokenization} and \textit{recommender training}. In the item tokenization stage, item metadata (e.g., name, description) is first encoded into pretrained semantic embeddings. These semantic embeddings are then tokenized by a pretrained tokenizer, which is often a vector quantizer~\cite{vqvae,rqvae} with a fixed vocabulary trained to reconstruct the original embeddings. The item token sequences, also known as semantic IDs, are cached and then used to train an LLM (e.g., T5~\cite{2020t5}) to predict the recommended next item.

\blue{The two-stage generative recommenders introduce an objective misalignment: the tokenizer learns to encode pretrained semantics and produces static token representations (semantic IDs) for items that remain fixed during recommender training, while the recommender is optimized for sequential behavior modeling. The two-stage objective misalignment prevents the static token representations from adapting to diverse recommendation contexts, and the recommender training erodes the pretrained semantic knowledge captured during tokenization.}
Specifically, we aim to address two limitations as follows:

\begin{figure}[t]
    \centering
    \includegraphics[width=\linewidth]{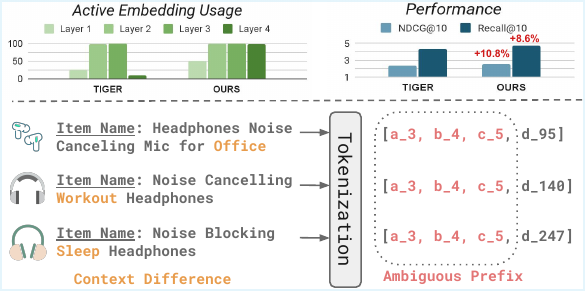}
    \caption{\textit{Suboptimal static tokenization} assigns identical prefix tokens to semantically distinct items (e.g., noise-canceling headphones for office use, workout, or sleep), leading to ambiguous representations that fail to reflect diverse user interaction contexts. }
    \label{fig:challenge}
\end{figure}

\textit{(C1)} \textbf{Suboptimal Static Tokenization:} Semantic IDs are produced by a pretrained tokenizer trained to reconstruct semantic embeddings rather than to optimize recommendation performance, leading to suboptimal tokenization for recommendation~\cite{tiger,letter}. 

\sigir{In particular, static tokenization imposes a representation bottleneck for recommenders that forces behaviorally distinct items to share identical representations for 75\% of the generation steps (prefix: $\texttt{[a\_3,b\_4,c\_5]}$), blinding the model to user intent until the final token generation. The consequent lack of discriminative input limits the attention mechanism, which fails to resolve user intent from structurally identical representations.}
As illustrated in Figure~\ref{fig:challenge}, items intended for different purposes (e.g., sleep, office, or workout use) receive identical prefix tokens, leading to \textit{prefix ambiguity}. While recent efforts have sought to address this issue by jointly optimizing the tokenizer and the recommender~\cite{etegrec, yin2025unleash}, such approaches require repeated re-tokenization throughout training, which introduces instability and increases computational overhead.



\textit{(C2)} \textbf{Discarded Pretrained Semantics:}  
During recommender training, the semantic knowledge encoded in pretrained embeddings is discarded after tokenization. While semantic IDs are trained to reconstruct pretrained semantic embeddings, the corresponding token embeddings are randomly initialized and trained solely on user interaction data. This two-stage design forfeits the semantics captured by the pretrained model (e.g., world knowledge in LLM embeddings)
. For example, token $\textit{a\_54}$ is pretrained to represent both the fruit and the technology brand meanings of “apple”. However, if $\textit{a\_54}$ appears predominantly in electronics-related contexts during training, its embedding captures only the tech brand, leading to misaligned recommendations for sparse items such as suggesting laptops to users seeking groceries.

While recent efforts have attempted to overcome these issues (\textit{C1} and \textit{C2}), they remain limited in important ways. For example, ETEGRec~\cite{etegrec} jointly optimizes the tokenizer and recommender in an end-to-end fashion with repeated re-tokenization during training, which leads to instability and results in suboptimal convergence (as illustrated in Figure~\ref{fig:convergence}). Other methods, such as LETTER~\cite{letter} and CoST~\cite{zhu2024costcontrastivequantizationbased}, \blue{aim to enhance the semantic knowledge preservation of item tokenization in the pretraining stage. However, the enhanced static tokenization still discards the fine-grained semantic information in pretrained embeddings in the recommender training stage, limiting the ability to leverage external knowledge. In contrast, our work provides a novel solution that avoids re-tokenization while retaining pretrained semantics, yielding a more stable yet expressive generative recommender framework.}

\blue{Specifically}, we propose \textbf{DECOR} (\underline{De}composed \underline{Co}ntextual Token \underline{R}epresentations), a novel framework that contextually adapts token embeddings with
pretrained semantic embeddings and collaborative signals in user interaction sequential modeling.
DECOR consists of two key components: \textit{Contextualized Token Composition} and \textit{Decomposed Embedding Fusion}. Specifically, \textit{Contextualized Token Composition} is a lightweight, dynamic interpretation mechanism applied during the recommender model's token generation to address the suboptimality of static tokenization (\textit{C1}). Instead of relying solely on the embeddings corresponding to static semantic IDs (as shown in Figure~\ref{fig:challenge}), each token embedding is contextually enhanced via a soft composition with token embeddings \blue{in a predefined candidate token set}
, conditioned on the user's interaction history. This contextual composition of tokens allows the LLM recommender to reinterpret token semantics in a context-dependent manner, compensating for the suboptimality of fixed token assignments. To preserve pretrained semantics (\textit{C2}), we retain the pretrained codebooks from the RQ-VAE tokenizer as frozen semantic embeddings and introduce separate, learnable collaborative embeddings. These two representations are fused through a light-weight fusion network guided by the recommendation loss, enabling the model to adaptively integrate pretrained semantic knowledge from the frozen codebook and interaction patterns learned during sequential modeling. 
Our contributions are listed as follows: 
\begin{itemize}
    \item We focus on two key limitations in existing generative recommenders: (i) the \textit{suboptimality of static semantic tokenization}, and (ii) the \textit{discarded pretrained semantics}. \blue{We are the first to explicitly analyze how static tokenization limits generative recommender’s ability to adapt item representations to diverse user context dynamics in generative recommendation.}
    \item We propose \textbf{DECOR} (Decomposed Contextual Token Representation Learning), a novel framework that fuses frozen pretrained semantic embeddings with learnable collaborative embeddings, and introduces a dynamic, context-aware token composition mechanism to dynamically interpret static input tokens.
    \item We conduct comprehensive experiments on three real-world datasets, demonstrating that DECOR consistently outperforms recent baselines. Notably, DECOR achieves faster convergence than the end-to-end tokenization–recommender joint training baseline while delivering higher accuracy with only moderate computational overhead.
\end{itemize}

\begin{figure*}[t]
    \centering
    \includegraphics[width=0.95\linewidth]{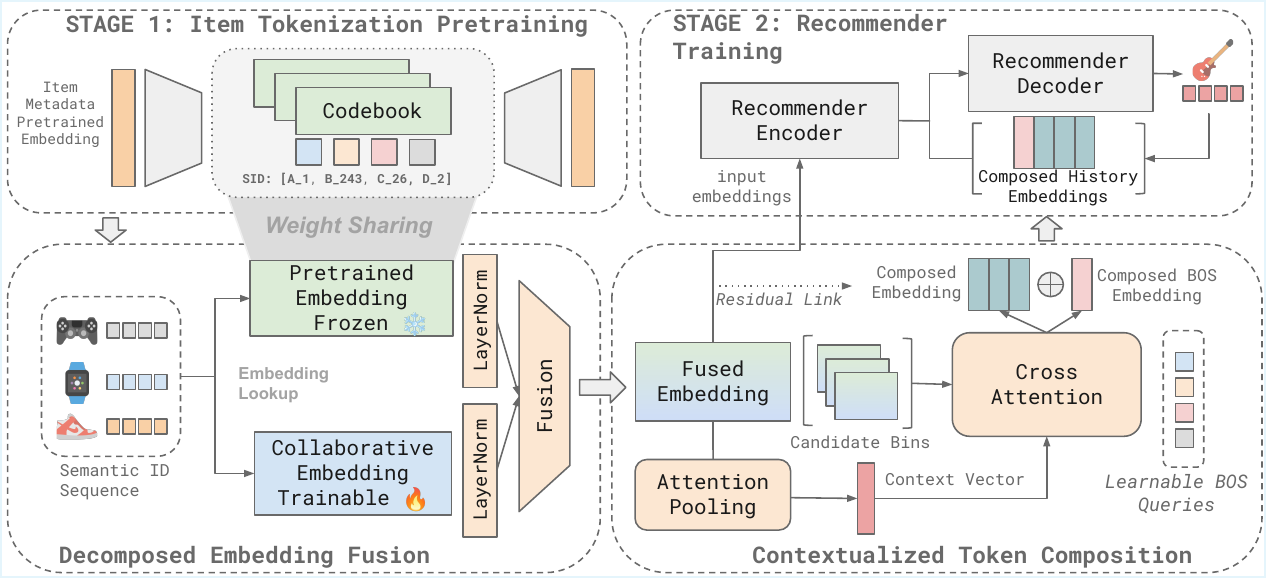}
    \caption{DECOR enhances generative recommendation via two components: decomposed embedding fusion integrates frozen pretrained embeddings and newly learned collaborative embeddings, while contextualized token composition dynamically refines token representations during autoregressive generation.}
    \label{fig:pipeline}
\end{figure*}

\section{Related Works}
\subsection{Generative Recommender}
In recent years, generative recommendation has emerged as a promising paradigm that formulates the sequential recommendation task as a sequence-to-sequence problem and directly generates the unique identifier of the next item~\cite{genrecsys_review}. Early work such as P5~\cite{p5} fine-tunes a pretrained language model (e.g., T5~\cite{2020t5}) to handle multiple recommendation tasks within a unified generative framework. TIGER~\cite{tiger} advances this paradigm by introducing a discrete semantic tokenization scheme based on item metadata (e.g., title, description), enabling the use of a pretrained T5 to autoregressively generate item token sequences. Building on TIGER, EAGER~\cite{eager} proposes a two-stream generation framework with a shared encoder and separate decoders to jointly capture user behavior and item content semantics. OneRec~\cite{onerec} further extends the generative approach by unifying retrieval and ranking into a single iterative generation process. Our work investigates the under-explored limitations of the tokenization pretraining-recommender training process in generative recommenders, addressing the suboptimality of static tokenization and the loss of pretrained semantics during training.


\subsection{Item Token Representation}
A key step in generative recommenders is to represent items as discrete tokens that can be consumed by language models. Early approaches fall into two broad categories: pseudo ID-based and text-based. To encode a large number of items, pseudo ID-based methods assign each item a unique token identifier without incorporating any semantic structure~\cite{chu2023leveraging,p5,index,wang2024enhanced}. Text-based methods, on the other hand, utilize item metadata (e.g., titles, descriptions) to construct natural language prompts representing input item sequences~\cite{bao2025bi,dai2023uncovering,li2023generative,liao2023llara,zhang2025recommendation,zhang2021language,liao2024llara}. While more expressive, text-based methods
incur high inference costs and may introduce hallucinated content~\cite{index}. To balance semantic fidelity and token efficiency, recent work introduces semantic indexing schemes that extract compact token sequences by quantizing pretrained text embeddings of item descriptions~\cite{tiger,letter,LCRec,zhu2024costcontrastivequantizationbased,general_VQ}. However, existing semantic indexers adopt static tokenization, where tokenized item representations remain fixed throughout training, limiting the model’s ability to 
adapt item representations
based on recommendation signals. 
To address the static tokenization limitation, ETEGRec~\cite{etegrec} jointly trains the item tokenizer and recommender model in an end-to-end fashion, though this coupling may introduce instability as token assignments evolve during training. ED$^2$~\cite{yin2025unleash} proposes a duo-index framework with a multi-grained token regulator and instruction tuning with user-level metadata, which may not be available in all scenarios. In contrast, our method enables stable yet contextually adaptive token interpretation during generation without relying on user-specific information. Moreover, a recent concurrent work, ActionPiece~\cite{hou2025actionpiece}, models context by applying BPE-based tokenization to group frequent item transitions into semantic units, enabling a standard Transformer to predict the next "action piece" via self-attention without requiring auxiliary parameter modules. Our work keeps the next-item prediction formulation with a context-enhanced item token representation.



\section{Method}
\subsection{Problem Formulation}
We consider the generative recommendation task under the sequential recommendation scenario. Formally, given the set of items $\mathcal{I}$ and a user's interaction sequence $\mathbf{S}^{u} = [i_{1}, i_{2}, \dots, i_{t-1}]\in \mathcal{I}$, the generative recommendation task is to predict the next item $i_{t} \in \mathcal{I}$.
Generative recommendation addresses this task through two key steps: item tokenization and autoregressive generation. Item tokenization maps each item $i \in \mathcal{I}$ into a discrete token sequence $\mathbf{c}_i = [c_{i,1}, c_{i,2}, \dots, c_{i,L}] \in \mathcal{C}$, where $L$ is the sequence length and $\mathcal{C}$ is predefined token set. The interaction sequence $\mathbf{S}^{u}$ is thereby transformed into a tokenized sequence $\mathbf{X}^{u} = [\mathbf{c}_{i_1}, \mathbf{c}_{i_2}, \dots, \mathbf{c}_{i_{t-1}}]$.
Given $\mathbf{X}^{u}$, the model autoregressively generates the token sequence $\mathbf{c}_{i_t}$ corresponding to the next target item $i_t$ by factorizing the conditional probability:
\begin{equation}
p(\mathbf{c}_{i_t}|\mathbf{X}^{u}) = \prod_{l=1}^{L} p(c_{i_t,l}|\mathbf{X}^{u}, c_{i_t,1}, \dots, c_{i_t,l-1}).
\end{equation}

\subsection{Semantic Indexer Pretraining}
\blue{As shown in the item tokenization pretraining stage of Figure~\ref{fig:pipeline}}, we use RQ-VAE~\cite{tiger} as our semantic indexer, which consists of a pair of MLP encoder-decoder, and a sequence of codebooks $\{\mathbf{C}_1, ..., \mathbf{C}_M\}$ where $M$ is the number of quantization levels for residual quantization.
Given a pretrained embedding of an item $\mathbf{x} \in \mathcal{R}^d$, \blue{the goal of a semantic indexer is to encode the pretrained semantics into a sequence of discrete tokens named semantic ID. Empirically, the length of the semantic ID obtained is $M+1$, where an additional token is appended to the output of the semantic indexer in order to resolve any duplicates. We denote the complete semantic ID length as $M_s = M+1$ in the subsequent discussions.}
\blue{The semantic indexing operates first by down-projecting the pretrained embedding vector to a latent vector $z_0$ by an encoder network.}
\begin{equation}
    \mathbf{z}_0 = \text{Encoder}(\mathbf{x}).
\end{equation}
\blue{Following the encoder projection, the hierarchical quantization is performed through $M$ residual steps:}
\begin{align}
    \mathbf{z}_m &= \mathbf{z}_{m-1} - q_l(\mathbf{z}_{m-1}), \quad m=1,...,M\\
    q_m(\mathbf{z}_{m-1}) &= \arg\min_{\mathbf{e} \in \mathbf{C}_m} \|\mathbf{z}_{m-1} - \mathbf{e}\|_2
\end{align}
where $q_l(\cdot)$ performs nearest-neighbor lookup with Euclidean distance in codebook $\mathbf{C}_m$. 

The quantized semantic ID of an item is the concatenation of code indices $\{i_1,..., i_{M_s}\}$ of the nearest vectors in codebooks from each level, with the final token for item collision handling. \blue{In particular, we adopt the collision handling scheme of~\cite{tiger}, which resolves items sharing the same first $M$ semantic IDs by appending incremental indices. For example, as shown in Figure~\ref{fig:pipeline}, an item may be assigned the semantic ID $[\texttt{A\_1}, \texttt{B\_243}, \texttt{C\_26}, \texttt{D\_2}]$, with the character prefixes included for clarity of demonstration.}

The training of RQ-VAE \blue{semantic indexer} is under the self-supervision of pretrained semantic embedding reconstruction. To perform reconstruction, \blue{each item's chosen codebook vectors are summed up, resulting in a codebook-estimated representation} $\tilde{\mathbf{r}} = \sum_{m=1}^M \mathbf{e}_{c_m}$. \blue{The estimated representation} $\tilde{\mathbf{r}}$ is then decoded back to semantic space where $\mathbf{e}_{c_m}$ is the corresponding codebook vector to semantic ID token $i_m$ in codebook $c_m$:
\begin{equation}
    \tilde{\mathbf{x}} = \text{Decoder}(\tilde{\mathbf{r}}).
\end{equation}
With the reconstructed representation $\tilde{\mathbf{x}}$, the whole RQ-VAE semantic indexer is optimized with $\mathcal{L}_\text{RQ}$ where:
\begin{align}
    \mathcal{L}_{\text{RQVAE}} &= \mathcal{L}_{\text{RECON}} + \mathcal{L}_{\text{RQ}} \\
    \mathcal{L}_{\text{RECON}} &= \|\mathbf{x} - \tilde{\mathbf{x}}\|_2^2 \quad \text{(reconstruction)} \\
    \mathcal{L}_{\text{RQ}} &= \sum_{m=1}^M \underbrace{\|\text{sg}[\mathbf{z}_{m-1}] - \mathbf{e}_{c_m}^m\|_2^2}_{\text{codebook update}} + \beta\underbrace{\|\mathbf{z}_{m-1} - \text{sg}[\mathbf{e}_{c_m}^m]\|_2^2}_{\text{commitment}}
\end{align}
where $\text{sg}[\cdot]$ denotes stop-gradient operation, and $\beta=0.25$ balances codebook learning~\cite{rqvae}. \blue{The reconstruction loss $\mathcal{L}_{\text{RECON}}$ ensures the reconstructed semantic embedding retains the original semantic meaning, and the commitment loss $\mathcal{L}_{\text{RQ}}$ encourages proximity between residual vectors and codebook embeddings.}

\subsection{Decomposed Embedding Fusion}\label{sec:fusion}

To retain the rich semantic information \textit{(C2)} in pretrained semantic embeddings, we introduce a \textit{Decomposed Embedding Fusion} module that dynamically fuses pretrained semantic and newly learned collaborative embedding representations. We treat the two modalities as complementary information channels and perform modality-aware fusion to adaptively integrate knowledge from both sources. 

Specifically, for pretrained semantics, we utilize the codebooks from a pretrained RQ-VAE tokenizer, which provides token-level representations learned to reconstruct pretrained embeddings while preserving the hierarchical structure of semantic IDs introduced by multi-stage tokenization~\cite{tiger}. Formally, we define the pretrained semantic embedding space for decomposed embedding fusion as:
\begin{equation}
E_{\text{pre}} = \{\mathcal{C}_i\}_{i=1}^{M} \in \mathbb{R}^{K\cdot M \times d},
\end{equation}
where each $\mathcal{C}_i \in \mathbb{R}^{K\times d}$ is a frozen codebook of $K$ embeddings for the $i$-th layer of an $M$-layer RQ-VAE semantic indexing scheme, and $d$ is the embedding dimension. To ensure compatibility with the downstream LLM recommender, we pretrain the tokenizer using codebooks with the same hidden size $d$ as the recommender. The pretrained semantic embedding $e_c$ for a token index $c_i$ is retrieved by a direct lookup from the corresponding codebook: $e_c = \mathcal{C}_i[c_i]$.
In parallel, we define the collaborative embedding space $E_{\text{collab}} \in \mathbb{R}^{K\cdot M \times d}$ as
a learnable embedding matrix trained from scratch via the autoregressive generation objective. Unlike the pretrained semantic embeddings guided by the reconstruction objective, collaborative embeddings are supervised purely based on user interaction sequences, allowing the model to encode sequential patterns such as co-occurrence and user preference dynamics. Notably, both $E_{\text{pre}}$ and $E_{\text{collab}}$ share the same dimensionality ($K \times d$), which enables seamless alignment and fusion in subsequent stages.

To bridge the modality gap between pretrained semantics and collaborative user interaction patterns~\cite{LCRec,mimoe}, we first project both embeddings into a shared latent space, followed by layer normalizations:
\begin{equation}
\hat{e}_{\text{pre}} = \text{LN}(W_{\text{pre}} e_{\text{pre}}), \quad
\hat{e}_{\text{collab}} = \text{LN}(W_{\text{collab}} e_{\text{collab}}),
\end{equation}
where $e_{\text{pre}}, e_{\text{collab}} \in \mathbb{R}^{d}$ are token embeddings from the pretrained and collaborative modalities respectively, $W_{\text{pre}}, W_{\text{collab}} \in \mathbb{R}^{d' \times d}$ are learnable projection matrices, and $\text{LN}$ denotes layer normalization.
We then concatenate the normalized embeddings and apply a fusion layer to map them back to the original latent space:
\begin{equation}
e_{\text{fused}} = W_{\text{fuse}} \left[ \hat{e}_{\text{pre}} \, \| \, \hat{e}_{\text{collab}} \right] \in \mathbb{R}^{d},
\end{equation}
where $W_{\text{fuse}} \in \mathbb{R}^{d \times 2d'}$ is a learnable fusion matrix, and $\|$ denotes vector concatenation. This fusion process enables the model to integrate both pretrained semantic and collaborative signals into a unified representation, aligning heterogeneous modalities while preserving their complementary strengths for downstream recommendation. During the forward pass, the decomposed embedding fusion module dynamically computes the fused embedding $\mathbf{e}_\text{fused}$ given the input token sequence $\mathbf{c}_i$ for each item $i$.

\subsection{Contextualized Token Composition}
\label{sec:CTC}
In two-stage generative recommendation paradigm, the static tokenization introduces a misalignment between an item’s static semantic ID and an item's dynamic latent representation learned for recommendation
Ideally, the semantic ID of an item should be dynamic and context-aware—capable of capturing pretrained semantic similarity for unseen items while also adapting to collaborative user interaction patterns learned. In particular, we identify the suboptimality of static tokenization \textit{(C1)} as illustrated in Figure~\ref{fig:challenge}, where multiple items are assigned identical tokens despite differing significantly in their usage contexts.

To better align tokenization to the recommendation objective, recent work introduces an end-to-end training framework where the tokenizer is jointly optimized with the LLM recommender in an iterative manner~\cite{etegrec}. However, the alternative optimizations of the tokenizer and recommender introduce training instability and lead to less-efficient training sessions. Instead of dynamically updating the semantic ID tokens of items, we propose to adapt token embeddings through composition with the embeddings of other tokens, enhancing token contextual expressiveness without iteratively performing re-tokenization.

To overcome the limitations of static item tokens, we introduce a contextualized token composition mechanism that refines token embeddings according to usage context. More formally, given a target item $i$ with a cached semantic ID sequence $\mathbf{c}_i = \{c_i^{(j)}\}_{j=1}^{M}$ and a historical context sequence $\mathbf{h}_{i}$, we compute the context-aware embedding for each token $c \in \mathbf{c}_i$ using a function $\Phi(c, u_c, \{ e_{c'} \}_{c' \in \mathcal{N}(c)})$, where $u_c$ is a context vector derived from the history and $\mathcal{N}(c)$ is a set of candidate composition tokens for $c$:
\begin{equation}
\label{eq:composition}
\tilde{e}_c = \Phi(c, u_c, \{ e_{c'} \}_{c' \in \mathcal{N}(c)}).
\end{equation}


\paragraph{Context Vector Computation.}
To obtain the context vector \( u_c \in \mathbb{R}^d \) used for generating a specific target token \( c \), we aggregate the fused embeddings of the historical context sequence \( \mathbf{h}_c = \{ h_1, h_2, \ldots, h_L \} \), where each \( h_\ell = f_{\text{fuse}}(c_\ell) \in \mathbb{R}^d \) denotes the fused embedding of token \( c_\ell \) obtained via the Decomposed Embedding Fusion module. Specifically, we apply an attention-based pooling mechanism to produce a summary of the context:
\begin{equation}
u_c = AttnPool(\mathbf{h_c}) =\mathrm{MLP}_\text{ctx} \left( \sum_{\ell=1}^{L} \alpha_\ell \cdot h_\ell \right),
\end{equation}
where the attention weights \( \{ \alpha_\ell \}_{\ell=1}^{L} \) are computed via:
\begin{align}
s_\ell &= \mathbf{w}^\top \tanh\left( \mathbf{W} h_\ell + \mathbf{b} \right)\\
\alpha_\ell &= \frac{\exp(s_\ell)}{\sum_{m=1}^{L} \exp(s_m)}.
\end{align}

Here, \( \mathbf{W} \in \mathbb{R}^{d' \times d} \), \( \mathbf{b} \in \mathbb{R}^{d'} \), and \( \mathbf{w} \in \mathbb{R}^{d'} \) are learnable parameters of the attention network, and \( \mathrm{MLP}_{\text{ctx}} \) is a multi-layer perceptron that transforms the weighted sum into the final context vector. The attention pooling allows the model to attend over the historical embeddings in a content-dependent manner and dynamically compute context \( u_c \).

\paragraph{Token Composition.}
To implement $\Phi$ in Equation~\ref{eq:composition}, we define a soft composition over a fixed set of candidate token embeddings $\{ e_{c'} \}_{c' \in \mathcal{N}(c)}$, where we choose the candidate set $\mathcal{N}(c)$ of token $c$ to be all tokens from same RQ-VAE codebook layer. Importantly, since $\mathcal{N}(c)$ includes all tokens within the same quantization layer, the token composition allows each token embedding to incorporate information from under-utilized or rarely selected codebook entries during pretraining. Our chosen candidate set enhances expressiveness by enabling the model to interpolate beyond the original static token assignments and leverage previously unused embedding capacity, effectively increasing the diversity of token representations. Empirically, the candidate set is chosen from the cached fused embeddings computed following Section~\ref{sec:fusion}. We then perform attention-based composition guided by the context vector $u$:
\begin{align}
\label{eq:composition_attention_weights}
\alpha_{c'} &= \frac{\exp\left( \langle W_q u, W_k e_{c'} \rangle \right)}{\sum_{c'' \in \mathcal{N}(c)} \exp\left( \langle W_q u, W_k e_{c''} \rangle \right)} \\
\tilde{e}_c &= \Phi_{\text{soft}}(c, u_c, \mathcal{N}(c)) = \sum_{c' \in \mathcal{N}(c)} \alpha_{c'} \cdot e_{c'}
\end{align}
where $W_q, W_k \in \mathbb{R}^{d \times d}$ are learnable projection matrices. The context-aware composed token embedding $\tilde{e}_c$ is fused with the original static embedding $e_c^{\text{static}}$ via a residual link:
\begin{equation}
e_{\text{final}} = \alpha \cdot \tilde{e}_c + (1 - \alpha) \cdot e_c^{\text{static}}, \quad \alpha \in [0, 1]
\end{equation}
Here, $\alpha$ is a tunable hyperparameter that controls the strength of context adaptation, which is fixed throughout the training. Smaller $\alpha$ values prioritize the static embedding, while larger values encourage reinterpretation based on the extracted user interaction context. Overall, contextualized token composition enables the model to flexibly incorporate collaborative signals into token representations at generation time, addressing \textit{suboptimal static tokenization} issue without modifying the tokenizer.

\subsubsection{Learnable BOS Embedding Composition}

For an RQ-VAE semantic indexing scheme, the first token in an item's semantic ID typically captures coarse-grained, high-level semantics—such as the item's broad category \blue{(e.g., ``headphones'' vs. ``books''). As a result, any ambiguity or mis-assignment at the first token decoding step propagates to the whole sequence, since autoregressive generation interprets subsequent tokens relative to that prefix. Consequently, accurately modeling the first token is critical: it serves as a global semantic anchor that conditions downstream decoding and strongly influences the interpretation of subsequent tokens. To refine the first token generation, we extend the token composition mechanism by introducing a set of $N$ learnable Beginning-of-Sequence (BOS) query vectors that adaptively refine the initial semantic anchor to align with the user’s context, denoted as $\mathcal{Q}_{\text{BOS}} \in \mathbb{R}^{N \times d}$, where $N$ is a hyperparameter controlling the number of BOS query vectors used.} The learnable BOS query vectors serve as latent representations of a set of candidate BOS tokens and allow the model to perform contextual composition for the BOS token embeddings, providing tailored BOS token embeddings based on the input token sequences. Specifically, for the generation of a target token $c$, the BOS token is composed with contextual composition function $\Phi$:
\begin{equation}
\hat{e}_{\text{BOS}} =  \Phi(e_\text{BOS}, u_c, \mathcal{Q}_{\text{BOS}} ),
\end{equation}
\noindent
\blue{where $e_\text{BOS}$ is a zero vector. Therefore $\hat{e}_{\text{BOS}}$ is simply composed by BOS query vectors in $\mathcal{Q}_{\text{BOS}}$. }The composed BOS embedding is used as the initial prefix for autoregressive generation of the item tokens, replacing the original BOS token. The learnable BOS queries form a unified scheme that ensures all generated token embeddings are dynamically adapted through contextualized composition. As a result, the model can better align the interpretation of each token with high-level semantic anchors (i.e., coarse-grained item semantics captured by the initial BOS composition), enabling more precise and coherent generation of semantic ID sequences.

\subsection{Complexity Analysis}
\textit{Decomposed Embedding Fusion} introduces a constant per-item cost of $\mathcal{O}(d^2)$ by projecting and combining token embeddings from pretrained and collaborative sources. \textit{Contextualized Token Composition} computes a context vector via attention pooling over $L$ historical tokens, with complexity $\mathcal{O}(L \cdot d^2)$. Each token composition attends to a fixed candidate set of size $K$, adding $\mathcal{O}(K \cdot d^2)$ per token. Since $d$ and $K$ are constants, the additional cost scales linearly with context length $L$ and remains \blue{controllable with a fixed candidate set} compared to the $\mathcal{O}(L^2 \cdot d)$ complexity of Transformer self-attention layers in the backbone recommender model. 

\blue{We empirically evaluate the inference cost of DECOR against the baseline (TIGER~\cite{tiger}) without \textit{Decomposed Embedding Fusion} or \textit{Contextualized Token Composition}. Our results in Table~\ref{tab:inference_efficiency} show that DECOR adds only \~0.85 ms of additional latency per sample at a batch size of 128 when considering the full candidate set for token composition, while consistently improving retrieval accuracy by 5--14\% over the base model.}

\subsection{DECOR Training}

As shown in Figure~\ref{fig:pipeline}, the proposed DECOR is integrated into every forward pass. We first apply decomposed embedding fusion to compute the encoder input embeddings by combining pretrained semantic embeddings with collaborative representations. During the autoregressive generation process, we replace the static embedding lookup with contextualized token composition, where each token embedding is dynamically adapted based on the generated context, which enables token embedding representations to evolve during training. To ensure semantic consistency, we reuse the fused vocabulary embeddings as the weights for the final prediction head. DECOR retains pretrained semantics and adapts to recommendation signals, effectively addressing both \textit{(C1)} suboptimal static tokenization and \textit{(C2)} discarded text semantics within a unified framework.




\begin{table}[t]
\centering
\caption{Statistics of the Datasets}
\label{tab:dataset_stats}
\begin{tabular}{lrrcc}
\toprule
Dataset & \#Users & \#Items & \#Interactions & Sparsity \\
\midrule
Scientific & 50,985 & 25,848 & 412,947 & 99.969\% \\
Instrument & 57,439 & 24,587 & 511,836 & 99.964\% \\
Game & 94,762 & 25,612 & 814,586 & 99.966\% \\
\bottomrule
\end{tabular}
\end{table}

\begin{table*}[t]
\centering
\setlength{\tabcolsep}{3.5pt}
\renewcommand{\arraystretch}{1.15}
\caption{Performance comparison on three datasets: Instrument, Scientific, and Game. Metrics include Recall@K (R@K) and NDCG@K (N@K). Bold indicates the best result per column. Superscript $^*$ indicates statistical significance at $p < 0.05$.}
\begin{tabular}{ll|cccc|cccc|cccc}
\toprule
\multirow{2}{*}{\textbf{Group}} & \multirow{2}{*}{\textbf{Method}} &
\multicolumn{4}{c|}{\textbf{Scientific}} &
\multicolumn{4}{c|}{\textbf{Instrument}} &
\multicolumn{4}{c}{\textbf{Game}} \\
& & R@5 & R@10 & N@5 & N@10 & R@5 & R@10 & N@5 & N@10 & R@5 & R@10 & N@5 & N@10 \\
\midrule
\multirow{6}{*}{Traditional} 
&Caser            & 0.0172 & 0.0281 & 0.0107 & 0.0142 & 0.0242 & 0.0392 & 0.0154 & 0.0202 & 0.0346 & 0.0567 & 0.0221 & 0.0291 \\
&GRU4Rec          & 0.0221 & 0.0353 & 0.0144 & 0.0186 & 0.0345 & 0.0537 & 0.0220 & 0.0281 & 0.0522 & 0.0831 & 0.0337 & 0.0436 \\
&SASRec            & 0.0256 & 0.0406 & 0.0147 & 0.0195 & 0.0341 & 0.0530 & 0.0217 & 0.0277 & 0.0517 & 0.0821 & 0.0329 & 0.0426 \\
&BERT4Rec          & 0.0180 & 0.0300 & 0.0113 & 0.0151 & 0.0305 & 0.0483 & 0.0196 & 0.0253 & 0.0453 & 0.0716 & 0.0294 & 0.0378 \\
&FDSA              & 0.0261 & 0.0391 & 0.0174 & 0.0216 & 0.0364 & 0.0557 & 0.0233 & 0.0295 & 0.0548 & 0.0857 & 0.0353 & 0.0453 \\
&S$^3$Rec          & 0.0253 & 0.0410 & 0.0172 & 0.0218 & 0.0340 & 0.0538 & 0.0218 & 0.0282 & 0.0533 & 0.0823 & 0.0351 & 0.0444 \\
\midrule
\multirow{8}{*}{\shortstack{Generative}}
&P5-SID            & 0.0155 & 0.0234 & 0.0103 & 0.0129 & 0.0319 & 0.0438 & 0.0237 & 0.0275 & 0.0480 & 0.0693 & 0.0333 & 0.0401 \\
&P5-CID            & 0.0192 & 0.0300 & 0.0123 & 0.0158 & 0.0352 & 0.0507 & 0.0234 & 0.0285 & 0.0497 & 0.0748 & 0.0343 & 0.0424 \\
&TIGER              & 0.0275 & 0.0431 & \underline{0.0181} & \underline{0.0231} & 0.0368 & 0.0574 & 0.0242 & 0.0308 & 0.0570 & 0.0895 & 0.0370 & 0.0471 \\
&LETTER         & \underline{0.0276} & \underline{0.0433} & 0.0179 & 0.0230 & 0.0372 & 0.0581 & 0.0243 & 0.0310 & 0.0576 & 0.0901 & 0.0373 & 0.0475 \\
&CoST           & 0.0270 & 0.0426 & 0.0180 & 0.0229 & 0.0366 & 0.0570 & 0.0242 & 0.0306 & 0.0569 & 0.0897 & 0.0379 & 0.0472 \\
& ETEGRec   & 0.0272 & \underline{0.0433} & 0.0173 & 0.0225 & \underline{0.0387} & \underline{0.0609} & \underline{0.0251} & \underline{0.0323} & \underline{0.0591} & \underline{0.0925} & \underline{0.0385} & \underline{0.0492} \\
& \textbf{DECOR} & \textbf{0.0309}* & \textbf{0.0469}* & \textbf{0.0206}* & \textbf{0.0257}* & \sigir{\textbf{0.0409}*} & \sigir{\textbf{0.0617}*} & \sigir{\textbf{0.0272}*} & \sigir{\textbf{0.0339}*} & \textbf{0.0610}* & \textbf{0.0944}* & \textbf{0.0400}* & \textbf{0.0507}*\\
\bottomrule
\end{tabular}
\label{tab:main_results}
\end{table*}
\section{Experiments}
In this section, we conduct experiments on three real-world datasets of the most updated Amazon Review dataset~\cite{hou2024bridging} to investigate the efficacy of our method. We follow the commonly adopted evaluation protocol and data preprocessing as prior works~\cite{tiger, letter}, and compare our method against a set of classic and state-of-the-art sequential recommendation baselines.

\subsection{Experimental Setting}
\subsubsection{Dataset}To validate the effectiveness of our method, we follow the commonly adopted evaluation protocol as prior works~\cite{tiger, letter}. Specifically, we conduct experiments on three subsets of the most updated Amazon Review dataset~\cite{hou2024bridging}. We apply the 5-core filter preprocessing, excluding items and users with fewer than five interaction records. After that, we construct user interaction sequences by aligning the items chronologically, with the maximum item sequence length set to 20. Our processed dataset statistics are in Table~\ref{tab:dataset_stats}. On all datasets, we evaluate all models using top-K Recall and NDCG with \( K = \{5, 10\} \). Following standard practice~\cite{tiger}, we adopt the leave-one-out strategy: for each user, the last interaction is used for testing, the second-last for validation, and the rest for training. We conduct a full-ranking evaluation over the entire candidate item set without sampling.

\subsubsection{Implementation Details}For our method, we first implement and reproduce the reported performance of TIGER~\cite{tiger}. Specifically, we use Sentence-T5~\footnote{https://huggingface.co/sentence-transformers/sentence-t5-base} as our text encoder for pretrained semantics and T5 as our generative recommender. Our experiments are carried out on a single NVIDIA Tesla A40 GPU. For reproducibility, we report the test set performance obtained with random seed 2025. To verify statistical significance, we conducted a paired t-test over five independent runs with seeds $\{2021, 2022, 2023, 2024, 2025\}$. In each run, the best model was selected based on validation set NDCG@10. For all experiments, we use a codebook size of 256 with 3 quantization levels that leads to 4-token semantic IDs.


\subsection{Baseline Models}We compare our methods against a set of classic ID-based sequential recommenders and LLM-based generative recommenders. \sigir{For generative recommenders, we re-run the official open-sourced code under identical experimental conditions to ensure fair comparison and to allow for training dynamics investigation. For other baselines that are originally trained under a different experimental setting, we adapt their official implementations to our experimental setting. And for baselines that are trained under the same experimental setting, we directly compare our performance with their reported performance in the paper.} The traditional baselines are:
\begin{itemize}
    \item \textbf{Caser~\cite{caser}} applies narrow convolutional filters vertically across rows to learn sequential patterns and horizontally across columns to detect co‑occurring latent features
    \item \textbf{SASRec~\cite{sasrec}} uses a stack of Transformer encoder layers with multi‐head self‐attention to model long‐range dependencies in user interaction sequences.
    \item \textbf{S$^3$Rec~\cite{s3rec}} enhances pre‐training by introducing four auxiliary self‐supervised tasks—masking attributes, predicting masked items, distinguishing subsequences, and contrasting full sequences—to maximize mutual information at multiple item granularities.
    \item \textbf{Bert4Rec~\cite{bert4rec}} employs a deep bidirectional Transformer encoder to learn rich, context‑aware item representations for sequential recommendation.
    \item \textbf{GRU4Rec~\cite{gru4rec}} encodes session‐based user behavior with a GRU network, then uses the final state to predict the next item.
\end{itemize}
The generative recommender baselines are:
\begin{itemize}
    \item \textbf{TIGER~\cite{tiger}} frames sequential recommendation as a generative retrieval task by quantizing item text embeddings via RQ‑VAE into a fixed vocabulary of semantic IDs and trains an LLM to autoregressively generate the next item's ID.
    \item \textbf{LETTER~\cite{letter}} optimizes an RQ‑VAE tokenizer by enforcing contrastive alignment and diversity regularization to learn hierarchical, collaborative, and diverse item tokens.
    \item \textbf{P5-CID~\cite{index}} performs spectral clustering on collaborative co‑occurrence graphs to group items, then uses the resulting cluster IDs as discrete tokens for generative recommendation.
    \item \textbf{P5-SID~\cite{index}} decomposes numeric item IDs into ordered subtokens (e.g., prefixes) so that frequently co‑occurring or sequentially adjacent items share subtoken patterns, improving locality in autoregressive generation.
    \item \textbf{CoST~\cite{zhu2024costcontrastivequantizationbased}} trains a quantization codebook with an InfoNCE‐style contrastive loss to map item embeddings into discrete semantic tokens that preserve both semantic similarity and neighborhood structure.
    \item \textbf{ETEGRec~\cite{etegrec}} jointly optimizes the tokenizer and the recommender model, with a set of alignment losses to improve tokenizer-recommender consistency.
\end{itemize}

\begin{table*}[t]
\centering
\setlength{\tabcolsep}{3.8pt}
\renewcommand{\arraystretch}{1.1}
\caption{
Ablation results of \textbf{DECOR} on three datasets. Each variant incrementally adds components: decomposed embedding fusion (\textit{DEF}), contextualized token composition (\textit{CTC}), and learnable BOS queries (\textit{BOS}). \cmark means the module is used, \xmark means not. Metrics include Recall@K (R@K) and NDCG@K (N@K). Bold indicates the best result per column. }
\begin{tabular}{ccc|cccc|cccc|cccc}
\toprule
\multicolumn{3}{c|}{\textbf{Modules}} &
\multicolumn{4}{c|}{\textbf{Scientific}} &
\multicolumn{4}{c|}{\textbf{Instrument}} &
\multicolumn{4}{c}{\textbf{Game}} \\
DEF & CTC & BOS & R@5 & R@10 & N@5 & N@10 & R@5 & R@10 & N@5 & N@10 & R@5 & R@10 & N@5 & N@10 \\
\midrule
\xmark & \xmark & \xmark & 0.0275 & 0.0431 & 0.0181 & 0.0231 & 0.0368 & 0.0574 & 0.0242 & 0.0308 & 0.0570 & 0.0895 & 0.0370 & 0.0471 \\
\xmark & \cmark & \xmark & 0.0292 & 0.0459 & 0.0193 & 0.0247 & 0.0385 & 0.0595 & 0.0261 & 0.0321 & 0.0599 & 0.0931 & 0.0394 & 0.0502 \\
\xmark & \cmark & \cmark & 0.0300 & 0.0462 & 0.0198 & 0.0248 & 0.0397 & 0.0605 & 0.0263 & 0.0329 & 0.0600 & 0.0932 & 0.0395 & 0.0500 \\
\cmark & \xmark & \xmark & 0.0294 & 0.0457 & 0.0192 & 0.0246 & 0.0382 & 0.0583 & 0.0254 & 0.0323 & 0.0602 & 0.0934 & 0.0390 & 0.0501 \\
\cmark & \cmark & \xmark & 0.0298 & 0.0465 & 0.0198 & 0.0250 & 0.0388 & 0.0598 & 0.0257 & 0.0324 & 0.0603 & 0.0932 & 0.0396 & 0.0500 \\
\cmark & \cmark & \cmark & \textbf{0.0301} & \textbf{0.0469} & \textbf{0.0201} & \textbf{0.0256} & \textbf{0.0409}* & \textbf{0.0617}* & \textbf{0.0272}* & \textbf{0.0339}* & \textbf{0.0610} & \textbf{0.0944} & \textbf{0.0400} & \textbf{0.0507} \\             
\bottomrule
\end{tabular}
\label{tab:ablation}
\end{table*}

\begin{figure*}[h]
  \centering
  \begin{subfigure}[b]{0.16\textwidth}
    \includegraphics[width=\linewidth]{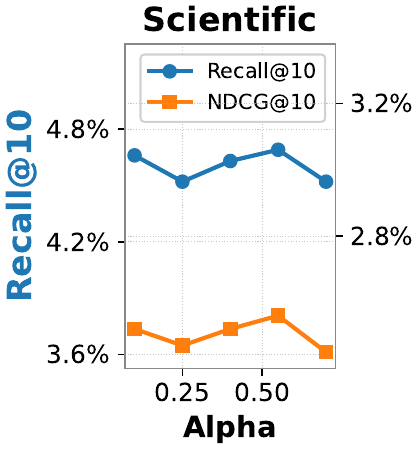}
  \end{subfigure}
  \begin{subfigure}[b]{0.16\textwidth}
    \includegraphics[width=\linewidth]{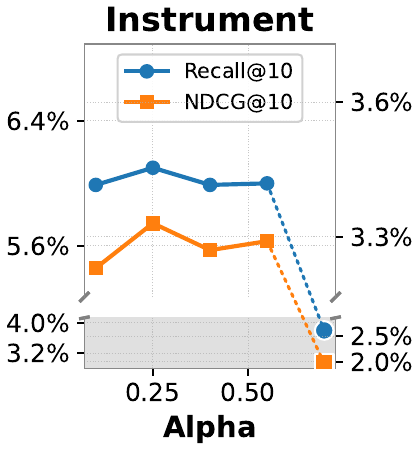}
  \end{subfigure}
  \begin{subfigure}[b]{0.16\textwidth}
    \includegraphics[width=\linewidth]{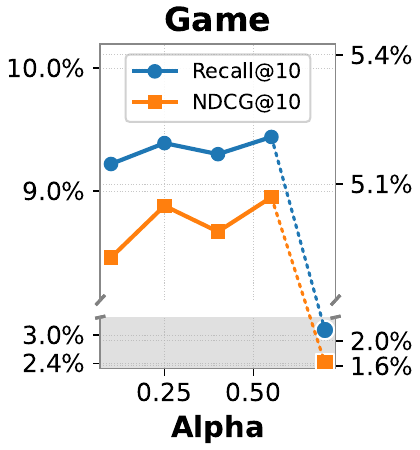}
  \end{subfigure}
  \begin{subfigure}[b]{0.16\textwidth}
    \includegraphics[width=\linewidth]{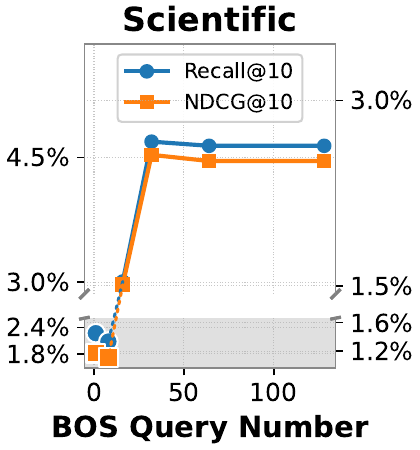}
  \end{subfigure}
  \begin{subfigure}[b]{0.16\textwidth}
    \includegraphics[width=\linewidth]{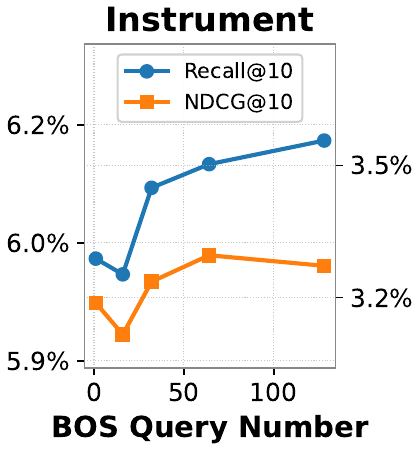}
  \end{subfigure}
  \begin{subfigure}[b]{0.16\textwidth}
    \includegraphics[width=\linewidth]{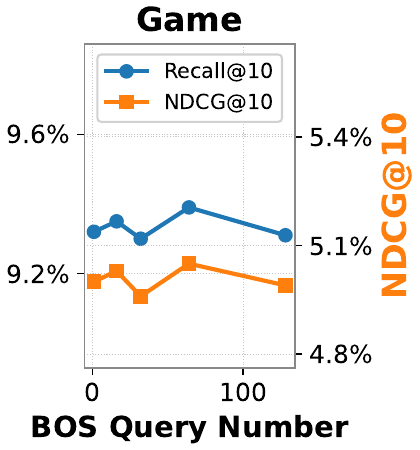}
  \end{subfigure}
  \caption{
    Parameter analysis of DECOR on contextual token composition weight $\alpha$ and BOS query number across all datasets. Shaded regions show performance collapse under extreme settings. We search the hyperparameters in the following region: $\alpha \in \{0.1,0.25, 0.4, 0.55, 0.7\}$ and BOS Query Number $\in \{8,16, 32, 64,128\}$
  }
  \label{fig:parameter_study}
\end{figure*}


\subsection{Overall Performance}
As shown in Table~\ref{tab:main_results}, DECOR consistently outperforms all baselines across metrics and datasets. Compared to traditional models such as SASRec and FDSA, it achieves substantial gains by effectively integrating pretrained semantics and collaborative signals. Among generative methods, DECOR not only surpasses static-tokenization approaches (e.g., TIGER, LETTER, CoST) but also outperforms ETEGRec, a dynamic baseline that jointly optimizes the tokenizer and recommender under alignment objectives. 


\sigir{Quantitatively, DECOR delivers substantial improvements across all domains, achieving relative gains of more than 14\% in NDCG@10 (Scientific), along with consistent improvements of 5\% and 3\% in Instruments and Games, respectively, over the strongest baselines. We specifically highlight the results on the Instrument dataset to illustrate the robustness of our approach. The instruments domain presents a unique challenge due to its reliance on specialized terminology (e.g., model numbers, tuning specs) rather than natural language descriptions. Consequently, generative baselines like TIGER encounter a performance bottleneck due to their reliance on static text priors. Similarly, the advanced dynamic tokenization baseline, ETEGRec, remains constrained as it attempts to align representations with these sparse pretrained semantic priors, failing to recover sufficient item distinctiveness. DECOR effectively circumvents this limitation. By decomposing the representation and explicitly fusing collaborative signals, our method compensates for the lack of semantic discriminability. The decomposed embedding fusion and contextualized decoding flow allow DECOR to capture fine-grained user preferences by adapting towards collaborative signals even when pretrained textual features are insufficient, securing a 5.0\% improvement in NDCG@10 over the strongest baseline.}



\subsection{Ablation Study}
Table~\ref{tab:ablation} presents an ablation analysis of DECOR, evaluating the contribution of contextualized token composition (CTC), learnable BOS queries (BOS), and decomposed embedding fusion with pretrained embeddings (DEF). \blue{Starting from the base model (TIGER), both adding contextualized token composition-only and decomposed embedding fusion yield similarly significant improvements across all datasets, confirming the value of preserving pretrained semantics and contextualized token refinement in recommender training.}
\blue{The best performance is achieved when both pretrained semantics and context-aware token composition are used together. Interestingly, we observe that adding BOS queries introduces performance gains when used jointly with decomposed embedding fusion, which suggests that BOS queries act as high-level semantic anchors that strengthen the contextual alignment between pretrained semantics and the composed token representations.} \sigir{While CTC requires a set of input embeddings to refine against the user context, there is no preceding item token that exists at step 0; the learnable BOS queries provide a parametrized context prefix before decoding, making the initial token generation better adapt to user history. Our ablation result demonstrates the efficacy of BOS (e.g., -2.43\% NDCG@10 Instruments).} Overall, the full DECOR consistently outperforms all ablations, demonstrating that each component contributes complementary benefits to the recommendation quality.

\subsection{Hyperparameter Sensitivity}
Figure~\ref{fig:parameter_study} examines the impact of two key hyperparameters in DECOR: the composition weight $\alpha$ and the BOS query number (defined in Section~\ref{sec:CTC}). For the visualization of each hyper-parameter, we fix the other to the best-performed value in NDCG@10.

\paragraph{Effect of Composition Weight $\alpha$.} DECOR is generally robust to a range of $\alpha$ values. Moderate values (e.g., $\alpha = 0.4$ to $0.55$) consistently yield the strongest performance across datasets, balancing contributions from the residual link and the context-aware composition. Note that even with $\alpha=0.1$, we observe consistent improvements compared to the ablated baseline (\textit{w/ Pretrained + Token Comp.} in Table~\ref{tab:ablation}). However, very high $\alpha$ values (e.g., $0.7$) result in sharp performance degradation, likely due to undertraining of individual token embeddings caused by excessive reliance on compositional signals from other tokens. The collapse in convergence is particularly pronounced on the Instrument and Game datasets, where recommendation performance drops significantly. We observe that Instrument and Game have larger interaction spaces, requiring token embeddings to stabilize early to generalize across diverse usage contexts. Excessive reliance on composition in such cases can delay embedding convergence and lead to training collapse.

\paragraph{Effect of BOS Query Number.} Under the best composition weight $\alpha$, increasing the number of BOS queries consistently enhances performance, particularly when increasing from 0 to 32. On the Scientific dataset, for example, NDCG@10 improves from 1.17\% to 2.56\%, indicating that BOS queries facilitate meaningful convergence by enabling the model to better capture diverse user preferences before generating the first token. However, beyond 32 (or 64 on the Game dataset), performance gains plateau, and larger values such as 128 yield no additional improvement, which we attribute to the model having already captured sufficient contextual information with a moderate number of BOS queries. Additional BOS queries likely provide redundant signals that do not contribute further to recommendation quality.

\subsection{Addressing Suboptimal Static Tokenization}

\begin{figure}[t]
  \centering
  \includegraphics[width=\linewidth]{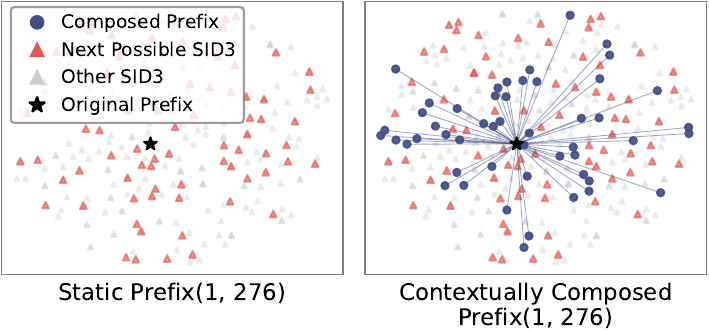}
  \caption{Case study for prefix ambiguity on the Scientific dataset. Compared to static tokenization (left), DECOR (right) produces prefix embeddings that are contextually adapted, enhancing expressiveness for disambiguation.}
  \label{fig:composed_visualization}
\end{figure}

To better understand whether our approach addresses suboptimal static tokenization, we present a case study on the Scientific dataset analyzing \textit{prefix ambiguity}. In Figure~\ref{fig:composed_visualization}, we present t-SNE visualizations of token embeddings before and after contextualized token composition. We observe that the static tokenization generates a single fixed embedding for the prefix $\texttt{(1,276)}$ due to deterministic embedding lookup, which lacks semantic coherence with valid next-token candidates (red triangles). In contrast, contextually composed prefix embeddings align more coherently with valid token candidates. The scattered composed prefix embeddings demonstrate that our method enhances prefix representations based on context, effectively mitigating the ambiguity of static tokenization. 

\blue{A well-known challenge in quantized representation learning for recommender systems is that many codebook entries remain inactive, leading to wasted capacity and limited representation diversity~\cite{letter}. For example, as shown in Table~\ref{tab:embedding_utilization}, only 25--28\% of active embeddings at the first quantization layer are utilized during tokenizer pretraining across all datasets.} In our experiments, we observe that DECOR improves the actively trained embedding coverage by involving inactive token embeddings in the token composition calculation, reaching 100\% on Instrument and Game, and 51.06\% on Scientific by filtering out tokens with below-uniform composition attention weights (Equation~\ref{eq:composition_attention_weights}). Notably, on Scientific, which has 20–50\% fewer interactions than the others (as shown in Table~\ref{tab:dataset_stats}), DECOR activates fewer additional token embeddings to accommodate the less diverse contextual modeling, thereby mitigating suboptimal tokenization with efficient representation usage.

\subsection{Comparison with Joint Tokenizer-Recommender Training}
\begin{figure}[t]
  \centering
  \includegraphics[width=\linewidth]{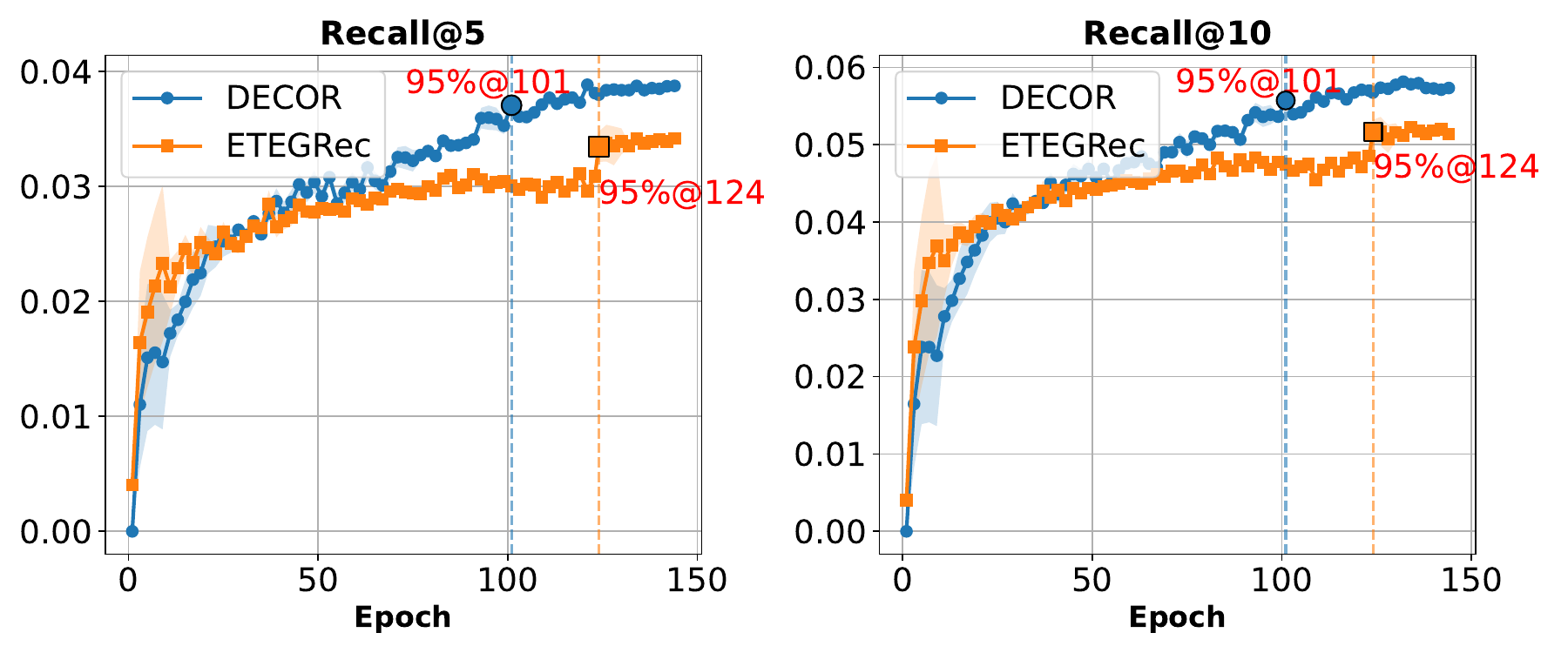}
  \caption{Convergence comparison of DECOR and ETEGRec on Recall@5 (left) and Recall@10 (right). DECOR not only achieves higher recall but also reaches 95\% of its peak performance substantially earlier (101 vs. 124 epochs), indicating faster and more stable convergence.}
  \label{fig:convergence}
\end{figure}

In Figure~\ref{fig:convergence}, we investigate the convergence behaviors on the Scientific dataset of DECOR and the tokenzier-recommender joint training baseline (ETEGRec~\cite{etegrec}) by comparing their validation performance throughout the training session. \sigir{Notably, direct runtime comparison is biased by implementations; therefore, we focus on sample efficiency as the robust metric. Figure 5 proves DECOR reaches 95\% peak performance 23 epochs faster than ETEGRec. Importantly, ETEGRec exhibits learning stagnation during epochs 75-110, whereas DECOR improves stably, validating DECOR's superior stability and sample efficiency compared to ETEGRec.} In particular, we exclude the epochs where ETEGRec optimizes its tokenizers and only report the validation results after every recommender update. We observe that while both models show improvements as the training progresses, DECOR consistently converges faster and to higher recall values. Specifically, DECOR reaches 95\% of its best performance at around epoch 101, while ETEGRec requires 124 epochs to achieve the same threshold. Our results indicate that DECOR not only achieves better final accuracy but also requires fewer training epochs to stabilize compared to ETEGRec. \sigir{In addition, we observe that the joint optimization of the tokenizer and recommender in ETEGRec adversely affects training stability. ETEGRec exhibits learning stagnation during epochs 75-110, whereas DECOR improves stably.} Once the tokenizer is fixed in the final training stage, however, ETEGRec exhibits a noticeable performance gain, confirming that iterative co-optimization constrains the recommender’s learning due to constantly changing item token representations and validating DECOR's superior stability and sample efficiency compared to ETEGRec.

\begin{table}[t]
\small
\centering
\caption{Comparison of active embedding usage at each quantization layer between the TIGER and DECOR.}
\label{tab:embedding_utilization}
\begin{tabular}{c|c|ccc}
\toprule
\multirow{2}{*}{\textbf{Dataset}} & \multirow{2}{*}{\textbf{Method}} & \multicolumn{3}{c}{\textbf{Code Embedding Utilization}} \\
& & \textbf{Layer-1} & \textbf{Layer-2} & \textbf{Layer-3} \\
\midrule
\multirow{2}{*}{\normalsize \textbf{Scientific}} 
& TIGER & 26.6\% & 99.07\% & 99.87\% \\
& Ours & \textbf{51.06\%} & \textbf{99.97\%} & \textbf{100.00\%} \\
\midrule
\multirow{2}{*}{\normalsize \textbf{Instrument}} 
& TIGER     & 27.97\% & 96.77\% & 100.00\% \\
& Ours & \textbf{100.00\%} & \textbf{100.00\%} & \textbf{100.00\%} \\
\midrule
\multirow{2}{*}{\normalsize \textbf{Game}} 
& TIGER    & 25.67\% & 99.61\% & 100.00\% \\
& Ours & \textbf{100.00\%} & \textbf{100.00\%} & \textbf{100.00\%} \\
\bottomrule
\end{tabular}
\end{table}

\subsection{Empirical Inference Cost Analysis}

\begin{table}[h]
\centering
\small
\setlength{\tabcolsep}{3.5pt}
\renewcommand{\arraystretch}{1.2}
\caption{Inference efficiency comparison using \textbf{Beam Search ($k=50$)}. 
Both methods demonstrate stable performance. DECOR introduces minimal latency overhead ($<$1ms) while maintaining $\approx$\textbf{82\%} of the baseline throughput.}
\begin{tabular}{c | cc | cc | c}
\toprule
\multirow{2}{*}{\textbf{Batch}} & \multicolumn{2}{c|}{\textbf{TIGER (Baseline)}} & \multicolumn{2}{c|}{\textbf{DECOR (Ours)}} & \textbf{Cost} \\
\cmidrule(lr){2-3} \cmidrule(lr){4-5}
 & \textbf{Lat.} (ms) & \textbf{Thr.} (s/s) & \textbf{Lat.} (ms) & \textbf{Thr.} (s/s) & \textbf{$\Delta$ ms} \\
\midrule
8   & 3.72 & 269 & 4.49 & 223 & +0.77 \\
16  & 3.68 & 272 & 4.49 & 223 & +0.81 \\
32  & 3.82 & 262 & 4.65 & 215 & +0.83 \\
64  & 3.83 & 261 & 4.67 & 214 & +0.84 \\
128 & 3.85 & 260 & 4.70 & 213 & +0.85 \\
\bottomrule
\end{tabular}
\label{tab:inference_efficiency}
\end{table}

\sigir{To empirically assess computational efficiency, we compare the generation inference cost with a beam search width of 50. As shown in Table~\ref{tab:inference_efficiency}, DECOR preserves the high efficiency of the underlying backbone. At a batch size of 128, the per-sample latency increases marginally from 3.85 ms (TIGER) to 4.70 ms (DECOR), representing a constant overhead of approximately 0.85 ms, which is primarily attributed to the fixed memory access patterns in the fusion module rather than computational complexity scaling of token composition. DECOR maintains a robust throughput of $>$210 samples/second, retaining approximately 82\% of the baseline's speed. Given the significant accuracy improvements of 5--14\% observed in Table~\ref{tab:main_results}, this sub-millisecond latency cost represents a highly favorable trade-off, confirming that DECOR is a cost-effective solution well-suited for real-time deployment.}



\section{Conclusion}
In this work, we address two limitations of existing generative recommenders: the suboptimal static tokenization and the discarded pretrained semantics. We propose DECOR, a unified framework that enhances token adaptability through contextualized token composition and preserves pretrained knowledge via decomposed embedding fusion. 
\blue{Experiments on three real-world datasets confirm that DECOR outperforms state-of-the-art baselines with higher accuracy and moderate computational cost, while ablation studies highlight DECOR's ability to bridge pretrained semantics and user–behavior dynamics in generative recommendation.}

\section{Acknowledgement}
This research is supported in part by the National Science Foundation under Grant No. CNS-2427070,  IIS-2331069,  IIS-2202481, IIS-2130263, CNS-2131622. The views and conclusions contained in this document are those of the authors and should not be interpreted as representing the official policies, either expressed or implied, of the U.S. Government. The U.S. Government is authorized to reproduce and distribute reprints for Government purposes notwithstanding any copyright notation here on.

\bibliographystyle{ACM-Reference-Format}
\bibliography{reference}



\end{document}